\documentclass[pdflatex,sn-basic,iicol]{sn-jnl}

\usepackage{graphicx}
\usepackage{amsmath,amssymb,amsfonts}
\usepackage{textcomp}
\usepackage{xcolor}
\usepackage{microtype}
\usepackage{booktabs}
\usepackage{tabularx}
\usepackage{array}

\begin{document}

\title[BPU-SpikeMM]{BPU-SpikeMM: Exact Matrix Multiplication via Spatio-Temporal Spike Encoding on Real Neuromorphic Hardware}

\author[1]{\fnm{Rongyuan} \sur{Guo}}\email{rongyuanguo@hnu.edu.cn}

\author*[1]{\fnm{Yufeng} \sur{Zhang}}\email{yufengzhang@hnu.edu.cn}

\author[1]{\fnm{Haotian} \sur{Wang}}

\author[2]{\fnm{Canming} \sur{Yao}}

\author[2]{\fnm{Yuxiang} \sur{Huan}}

\author[1]{\fnm{Wangdong} \sur{Yang}}

\author[1]{\fnm{Kenli} \sur{Li}}

\affil[1]{\orgdiv{College of Computer Science and Electronic Engineering},
  \orgname{Hunan University},
  \orgaddress{\street{Lushan South Road, Yuelu District},
    \city{Changsha}, \postcode{410082}, \state{Hunan}, \country{China}}}

\affil[2]{\orgname{Guangdong Institute of Intelligence Science and Technology},
  \orgaddress{\street{Building 6, 2515 North Huandao Road,
      Hengqin Guangdong-Macao In-Depth Cooperation Zone},
    \city{Zhuhai}, \postcode{519031}, \state{Guangdong}, \country{China}}}

\abstract{
Matrix multiplication is the dominant computational kernel in deep learning, yet its growing scale strains conventional von Neumann hardware in energy efficiency and bandwidth. Neuromorphic architectures offer a promising alternative, but mathematical equivalence between matrix multiplication and spiking computation does not guarantee direct execution on real neuromorphic hardware, as the specific hardware paradigm may still require additional data recovery, bit-weight reconstruction, or state readout after neuromorphic execution.

This paper presents \textbf{BPU-SpikeMM}
(\textbf{B}rain \textbf{P}rocessing \textbf{U}nit-based \textbf{Sp}ike
\textbf{M}atrix \textbf{M}ultiplication), an exact spatio-temporal spike
encoding framework for matrix multiplication on real neuromorphic hardware.
We prove the exact equivalence of spike-encoded matrix multiplication under
the BPU execution paradigm, distinguish mathematical equivalence from native
direct-readout equivalence on BPU hardware, and build a spatio-temporal
offloading pipeline supporting multiple encoding schemes. 
Furthermore, we extend the framework to a multi-dimensional encoding design space and evaluate the applicability of different encodings on real BPU hardware in terms of correctness, temporal resource cost, spatial resource cost, and event-driven metrics. The results confirm native direct-readout equivalence across all evaluated encoding schemes and show that their applicability to BPU execution depends on joint trade-offs among temporal cost, spatial resource consumption, and event activity.
}

\keywords{Neuromorphic Computing, Spike Encoding, Matrix Multiplication, Spatio-Temporal Parallelism, Hardware Resource Analysis}

\maketitle

\section{Introduction}

\textbf{Computing Demand and Neuromorphic Opportunity.} Matrix multiplication is the core operator in deep learning training,
inference, and scientific computing. Its growing scale imposes significant
pressure on computation, memory bandwidth, and energy
consumption~\citep{lei2023toward,demirkiran2024blueprint}. Although
conventional hardware continuously optimizes matrix multiplication
performance, data movement and power consumption remain the primary
bottlenecks. Consequently, the execution efficiency of matrix multiplication
often directly determines the throughput and energy-efficiency ceiling of
intelligent computing systems. However, current neuromorphic platforms generally do not match the raw computational performance of highly optimized conventional hardware. Nevertheless, neuromorphic hardware offers a promising pathway toward energy-efficient matrix computation: its event-driven computation, sparse spike communication, near-memory synaptic accumulation, and neuronal state evolution mechanisms provide a distinct hardware basis for low-power execution.~\citep{kudithipudi2025scale,shrestha2024efficient,kim2023snpu}.

Against this background, matrix multiplication on real neuromorphic hardware faces \textbf{two key challenges.}

\textbf{Challenge~1.} The mathematical equivalence between matrix multiplication and spike accumulation does not imply that exact direct-readout execution on real neuromorphic hardware is straightforward~\citep{date2023encoding,schuman2021sparse,
parekh2018threshold,pedersen2024nir}. For example, binary bit-plane encoding can represent multiplication exactly in mathematics, but a purely temporal weighted form is difficult to execute natively when the hardware cannot vary synaptic weights across time steps. Instead, weighted terms can be mapped to spatial synaptic groups with pre-deployed scaled weights, so that the final neuron state can be directly read out as the exact matrix multiplication result. This direct-readout property is particularly suitable for inference, where weights are deployed once and reused across many inputs.

\textbf{Challenge~2.} The advantages and applicability of different exact encoding schemes on neuromorphic hardware cannot be assessed by any single factor; rather, they depend on the interplay of
encoding sparsity, time-step overhead, spatial replication, synaptic weight
range, neuronal state readout, and event-driven metrics~\citep{bhattacharjee2024snns,
ottati2023spike}. A unified framework for comparing different exact encoding
schemes is still lacking. For example, the value~15 requires four active
terms in binary encoding, but only two in CSD encoding~(\(16-1\)) or
Fibonacci encoding~(\(13+2\)); these terms can also be unfolded into spatially
parallel channels to reduce temporal cost. Therefore, exact spike-based matrix
multiplication should be evaluated jointly from temporal, spatial,
weight-scaling, readout, and event-driven perspectives. Such a unified
spatio-temporal analysis is essential for guiding practical deployment on real
neuromorphic hardware and motivating hardware--software co-design.

To tackle these two challenges, this paper proposes \textbf{BPU-SpikeMM}, targeting real Brain Processing Unit (BPU)
hardware to systematically transform matrix multiplication into spike-domain
execution. We first establish, under real hardware execution constraints,
the exact equivalence conditions between matrix multiplication and spike
accumulation. We then construct a unified spatio-temporal parallelism mapping
framework based on this equivalence, supporting eight exact-equivalent encoding schemes with extensibility to additional encodings. Finally, we characterize the encoding design space from multiple dimensions,
including correctness, temporal resource cost, spatial resource cost, and
event-driven metrics, providing a basis for the engineering deployment of
matrix multiplication on neuromorphic hardware.

The main contributions of this paper are as follows:

\begin{itemize}
        \item We establish exact direct-readout equivalence for spike-based matrix multiplication based on the BPU hardware paradigm, beyond purely mathematical equivalence.

  \item We construct the BPU-SpikeMM spatio-temporal parallelism mapping
        framework, which represents matrix multiplication uniformly as a
        composite process of spike events, synaptic weights, and neuronal
        state readout.

    \item We validate BPU-native direct-readout exactness across eight exact encoding schemes on real BPU hardware and systematically characterize their trade-offs using temporal, spatial, and event-activity metrics.
  \item To the best of our knowledge, BPU-SpikeMM is the first work
to formally define BPU-native direct-readout equivalence for spike-domain matrix multiplication and to use this definition to systematically analyze and evaluate multiple encoding schemes on real neuromorphic hardware. The implementation will be made publicly available to support reproducibility and further research.
\end{itemize}

\section{Background and Related Work}

Matrix multiplication is a fundamental computational kernel whose
efficiency depends on both software/mathematical formulation and the
underlying hardware execution paradigm. Conventional hardware has developed multiple acceleration
approaches, including GPU/TPU tensor compute units, FPGA dataflow
acceleration, ASIC systolic arrays, sparse/dense matrix-specific
accelerators, and analog/in-memory computing
architectures~\citep{jouppi2023tpuv4,taka2024gemm,qin2020sigma,
yang2024trapezoid,lu2024aces,demirkiran2024blueprint,legallo2024aimc}.
These methods improve throughput and energy efficiency through
large-scale parallelism, data reuse, reduced-precision computation, or
near-memory computing, yet remain constrained by memory bandwidth, data
movement, irregular sparse access, analog-to-digital conversion, and
hardware resource scaling. Thus, efficient matrix computation must be considered from both the
software/mathematical level and the underlying hardware execution paradigm.

Combining software/mathematical formulation with the underlying hardware
execution paradigm provides another route to rethink matrix computation
beyond conventional binary computing. Unary and temporal computing represent
numerical values as unary bit streams or temporal pulse widths, enabling
computation with simpler logic. Representative works include \mbox{uGEMM},
which supports rate-coded and temporal-coded bit streams for matrix
multiplication~\citep{wu2020ugemm}, and tubGEMM, which uses a
temporal-unary-binary hybrid representation to perform exact matrix
multiplication and exploit numerical sparsity~\citep{vellaisamy2023tubgemm}.
Unary coding has also been used for ReRAM synaptic weight representation to
mitigate device variation~\citep{ma2020gounary}. These works show that
alternative representations can benefit matrix computation, but their
effectiveness must be evaluated under specific hardware paradigms rather
than mathematical representation alone.

Neuromorphic hardware further extends this idea by using event-driven
spike communication, near-memory synaptic accumulation, and neuronal state
dynamics as the execution substrate. Real neuromorphic platforms such as
Loihi~2 have demonstrated energy-efficiency potential in audio/video
processing, sequential tasks, and large-scale brain-inspired
systems~\citep{shrestha2024efficient,kudithipudi2025scale}. Beyond SNN
inference, recent studies have also explored numerical computation on
neuromorphic hardware, including spike-based encoding of integers and
rational numbers~\citep{date2023encoding} and sparse finite-element linear
solvers on Loihi~2~\citep{theilman2025fem}. 

Neuromorphic and threshold-circuit studies have also investigated
matrix operations through spike-based or neuron-like computational
models. Schuman et al. mapped sparse binary matrix--vector multiplication onto
neuromorphic computers, illustrating an explicit construction based on
binary inputs, synaptic weights, and neuronal accumulation
\citep{schuman2021sparse}. Parekh et al. analyzed matrix multiplication
using threshold circuits and established theoretical constructions and
complexity bounds for fully spatial representations
\citep{parekh2018threshold}. These studies provide important
foundations for representing matrix operations through neuromorphic or
threshold-based computation. 

Recent work also shows that the
advantages of SNN or neuromorphic systems must be evaluated jointly across
hardware platforms, workloads, encoding sparsity, measurement methodology,
and system overhead~\citep{bhattacharjee2024snns,yik2025neurobench,
ottati2023spike}. However, these works do not provide a unified hardware-aware
formulation of how multiple exact matrix-multiplication encodings can
achieve native direct-readout execution under a common real BPU-class
hardware paradigm.

In contrast, this paper studies how matrix multiplication can be
systematically transformed into spike-domain execution under the BPU
hardware paradigm while satisfying native direct-readout equivalence.
Rather than introducing a new numerical encoding, BPU-SpikeMM provides
a unified hardware-aware formulation for analyzing multiple exact
encoding schemes under common execution constraints. The framework
jointly characterizes correctness, temporal cost, spatial resource
consumption, and event activity, thereby exposing the distinct
trade-offs among different encoding schemes without assuming that any
single scheme is universally optimal.

\section{Hardware-Aware Encoding Equivalence for Matrix Multiplication}

This section establishes exact encoding equivalence for matrix multiplication
under the BPU hardware paradigm. Sections~\ref{AA} and~\ref{BB} derive and
prove the exact equivalence condition by combining mathematical equivalence
with the real BPU execution paradigm. Sections~\ref{CC} and~\ref{DD} then
provide five concrete instantiations to help distinguish mathematical
equivalence from exact equivalence (native direct-readout equivalence under
the BPU hardware paradigm). Section~\ref{EE} summarizes the exact equivalence results and spatio-temporal characteristics, motivating the mapping framework in the next section.

\subsection{Problem Formulation and BPU Execution Model}\label{AA}

We consider the standard matrix multiplication between an activation matrix
$\mathbf{A}\in\mathbb{R}^{M\times K}$ and a weight matrix
$\mathbf{B}\in\mathbb{R}^{K\times N}$:
\begin{equation}
    \mathbf{Y}=\mathbf{A}\mathbf{B}, \qquad
    \mathbf{Y}\in\mathbb{R}^{M\times N}.
\end{equation}
For the $m$-th input row, the $j$-th output element is computed as
\begin{equation}
    y_{m,j}=\sum_{i=1}^{K} a_{m,i} b_{i,j},
    \label{eq:dense_matmul}
\end{equation}

\begin{figure*}[t]
    \centering
\includegraphics[width=\textwidth]{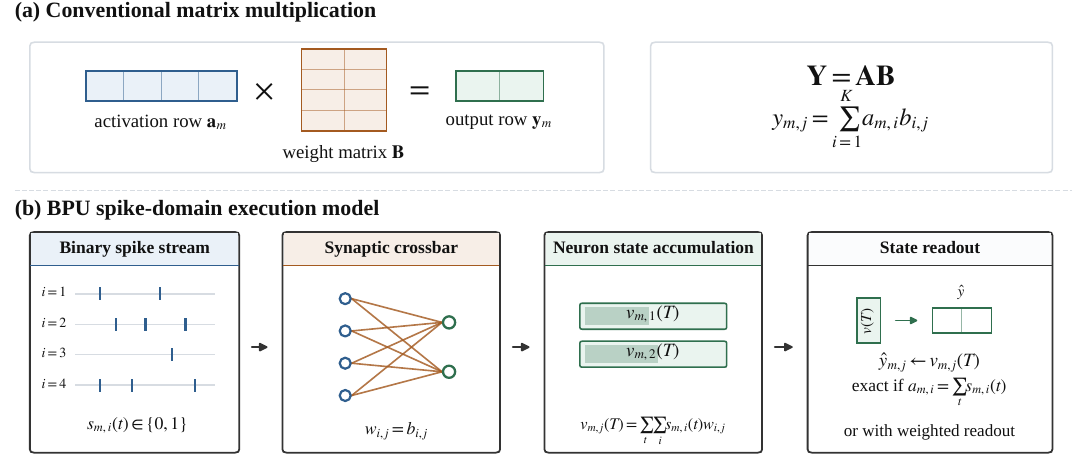}
\caption{Correspondence between conventional matrix multiplication and
BPU spike-domain execution}
    \label{fig:bpu_execution_model}
\end{figure*}

As illustrated in Fig.~\ref{fig:bpu_execution_model}(a), $a_{m,i}$ denotes the $i$-th activation and $b_{i,j}$ denotes the
corresponding weight. This row-wise formulation naturally matches the
execution style of a fully connected neuromorphic layer, where the input
channels correspond to the $K$ activation dimensions and the output neurons
correspond to the $N$ output dimensions.

On the BPU, numerical activations are not directly injected as analog values.
Instead, each activation $a_{m,i}$ is represented by a binary spike train
$s_{m,i}(t)\in\{0,1\}$ over discrete time steps $t=1,\ldots,T$. The weight
matrix is mapped to static synaptic connections:
\begin{equation}
    w_{i,j}=b_{i,j}.
\end{equation}
When a spike arrives at input channel $i$, the synapse contributes $w_{i,j}$
to the state of output neuron $j$. In a general discrete-time form, the output
neuron state can be written as
\begin{equation}
    v_{m,j}(t+1)
    =
    \lambda v_{m,j}(t)
    +
    \sum_{i=1}^{K} s_{m,i}(t)w_{i,j},
    \label{eq:bpu_state_general}
\end{equation}
where $\lambda$ denotes the state retention factor determined by the neuron
dynamics. As illustrated in Fig.~\ref{fig:bpu_execution_model}(b), for exact matrix multiplication, we focus on the linear accumulation
regime, in which leakage and reset effects are either disabled, avoided by
state readout before firing, or compensated by configuration. Under this
regime, the final output state becomes
\begin{equation}
    v_{m,j}(T)
    =
    \sum_{t=1}^{T}\sum_{i=1}^{K} s_{m,i}(t)w_{i,j}.
    \label{eq:bpu_linear_accumulation}
\end{equation}

The BPU result is obtained by reading out the final neuron state:
\begin{equation}
    \hat{y}_{m,j}=\mathcal{R}\left(v_{m,j}(T)\right),
    \label{eq:bpu_readout}
\end{equation}
where $\mathcal{R}(\cdot)$ denotes the state readout and possible scaling
operation. Therefore, the key problem is to design a spike representation
$s_{m,i}(t)$ and, when necessary, a readout or weight-scaling mechanism such
that $\hat{y}_{m,j}=y_{m,j}$ for all output dimensions. The following
subsections derive the general equivalence condition and instantiate it with
native spike-count and weighted bit-plane encodings.

\subsection{General Encoding Equivalence Condition}\label{BB}

Based on the BPU linear accumulation model, we now derive a general condition
for exact equivalence between matrix multiplication and spike-domain
computation. The key idea is to represent each numerical activation as a
weighted sum of binary spike events. Specifically, for each activation
$a_{m,i}$, we consider an exact spike representation of the form
\begin{equation}
    a_{m,i} = \sum_{\ell=1}^{L} q_{\ell} s_{m,i}^{(\ell)},
    \qquad
    s_{m,i}^{(\ell)}\in\{0,1\},
    \label{eq:general_spike_encoding}
\end{equation}
where $s_{m,i}^{(\ell)}$ denotes a binary spike variable and $q_{\ell}$
denotes the contribution weight associated with the $\ell$-th spike position,
time step, or spatial channel.

Substituting \eqref{eq:general_spike_encoding} into the dense matrix
multiplication in \eqref{eq:dense_matmul}, we have

\begin{equation}
\begin{aligned}
y_{m,j}
&= \sum_{i=1}^{K} a_{m,i} b_{i,j}
 = \sum_{i=1}^{K}\left(\sum_{\ell=1}^{L} q_{\ell}s_{m,i}^{(\ell)}\right)b_{i,j} \\
&= \sum_{\ell=1}^{L} q_{\ell}
   \left(\sum_{i=1}^{K} s_{m,i}^{(\ell)} b_{i,j}\right).
\end{aligned}
\label{eq:general_equivalence_derivation}
\end{equation}

The inner summation in \eqref{eq:general_equivalence_derivation} has the same
form as a BPU synaptic accumulation step. If the weight matrix is mapped to
the synapses as $w_{i,j}=b_{i,j}$, the partial spike-domain contribution can
be written as
\begin{equation}
    u_{m,j}^{(\ell)}
    =
    \sum_{i=1}^{K} s_{m,i}^{(\ell)} w_{i,j}.
    \label{eq:partial_spike_contribution}
\end{equation}
Therefore, exact equivalence is obtained if the final readout satisfies
\begin{equation}
    \hat{y}_{m,j}
    =
    \sum_{\ell=1}^{L} q_{\ell} u_{m,j}^{(\ell)}
    =
    y_{m,j}.
    \label{eq:general_exact_condition}
\end{equation}

Equation \eqref{eq:general_exact_condition} gives the general equivalence
condition. It separates the problem into two parts: the numerical activation
must first be represented exactly by binary spike variables as in
\eqref{eq:general_spike_encoding}, and the corresponding contribution weights
$q_{\ell}$ must then be realized either by the hardware execution path or by
the readout mechanism. When $q_{\ell}=1$ for all $\ell$, the computation
reduces to native spike-count accumulation. When $q_{\ell}$ takes different
values, such as the bit weights in binary coding, the scaling factors must be
implemented through temporal readout weighting, spatial weight scaling, or
other hardware-supported mechanisms.

This condition unifies different exact encodings: spike-count encoding is the
native unit-weight case, while bit-plane and other weighted encodings require
their contribution weights to be realized under the BPU execution constraints.

\subsection{Exact Equivalence of Spike-Count Encoding}\label{CC}

We first consider the most native case for BPU execution: spike-count
encoding. Here all spike events have unit contribution weight, i.e.,
$q_{\ell}=1$ in \eqref{eq:general_spike_encoding}. For a non-negative integer
activation,
\begin{equation}
    a_{m,i}=\sum_{\ell=1}^{L}s_{m,i}^{(\ell)},\qquad
    s_{m,i}^{(\ell)}\in\{0,1\}.
    \label{eq:spike_count_encoding}
\end{equation}
Thus, the value is represented only by the number of spikes, matching the
native BPU rule that each incoming spike triggers the same synaptic
contribution $w_{i,j}=b_{i,j}$.

Spike-count encoding can be realized temporally, spatially, or jointly in
both dimensions:
\begin{align}
    a_{m,i} &= \sum_{t=1}^{T}s_{m,i}(t), \label{eq:temporal_unary}\\
    a_{m,i} &= \sum_{r=1}^{R}s_{m,i,r},\quad
               w_{(i,r),j}=b_{i,j}, \label{eq:spatial_unary}\\
    a_{m,i} &= \sum_{t=1}^{T}\sum_{r=1}^{R}s_{m,i,r}(t),\quad
               w_{(i,r),j}=b_{i,j}. \label{eq:spatio_temporal_unary}
\end{align}
Substituting any of these representations into the linear BPU accumulation
model yields
\begin{equation}
    v_{m,j}=\sum_{i=1}^{K}a_{m,i}b_{i,j}=y_{m,j},
    \label{eq:spike_count_equivalence}
\end{equation}
where $v_{m,j}$ denotes the final accumulated state after the required time
steps. Therefore, spike-count encoding achieves native direct-readout
equivalence: no dynamic weight scaling or weighted readout is required, and
the final BPU neuron state directly corresponds to the matrix multiplication
result.

\begin{figure*}[t]
    \centering
\includegraphics[width=\textwidth]{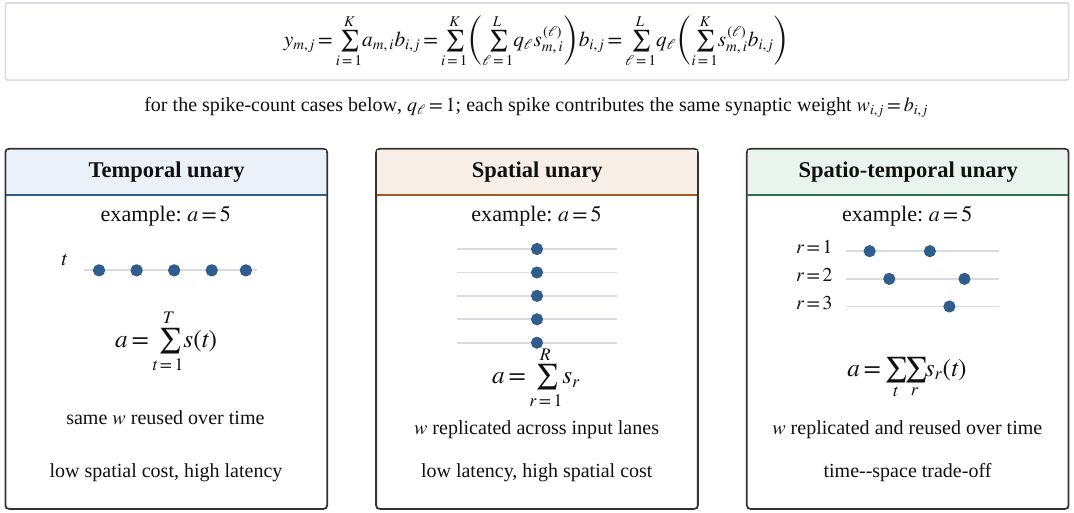}
\caption{Spike-count encoding variants for native and spatio-temporal BPU equivalence}
    \label{fig:spike_count_encoding_variants}
\end{figure*}

Fig.~\ref{fig:spike_count_encoding_variants} illustrates temporal, spatial,
and spatio-temporal spike-count mappings. Temporal unary coding reuses the
same synaptic matrix over time, spatial unary coding replicates input lanes
with identical synaptic weights, and spatio-temporal unary coding trades
spatial replication for additional time steps. Since all spikes carry the same
unit contribution weight, these mappings are directly compatible with the BPU
execution model.

\begin{figure*}[t]
    \centering
\includegraphics[width=\textwidth]{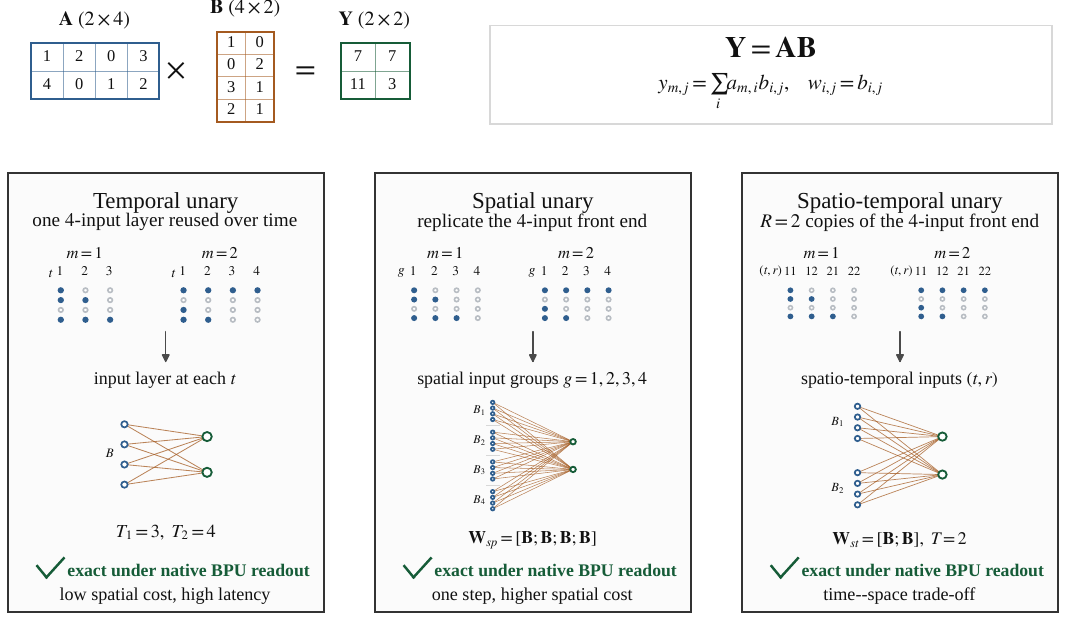}
\caption{Performance of unary encoding in exact matrix multiplication equivalence}
    \label{fig:matrix_spike_count_mapping}
\end{figure*}

To make the mapping concrete, Fig.~\ref{fig:matrix_spike_count_mapping}
shows a $2\times4$ by $4\times2$ matrix multiplication example. The weight
matrix $\mathbf{B}$ is mapped to a $4$-input and $2$-output BPU layer. Temporal
unary coding reuses this layer over time, spatial unary coding replicates the
complete $4$-input front end across spatial groups, and spatio-temporal unary
coding moves only part of the temporal expansion to space. This example also
highlights that spatial expansion should preserve the original input-channel
structure: each replicated group corresponds to one complete $4\times2$
synaptic block rather than an unstructured set of spike lanes.

\subsection{Exact Equivalence of Bit-Plane Encoding}\label{DD}

Bit-plane encoding is a representative weighted spike encoding. For a
non-negative integer activation $a_{m,i}$ with $K_b$ bits, the $k$-th
bit-plane spike is defined as
\begin{equation}
s_{m,i}^{(k)} =
\left\lfloor \frac{a_{m,i}}{2^k} \right\rfloor \bmod 2,
\quad k=0,\ldots,K_b-1 ,
\end{equation}
and the activation can be reconstructed as
\begin{equation}
a_{m,i} = \sum_{k=0}^{K_b-1}2^k s_{m,i}^{(k)} .
\label{eq:bit_plane_activation}
\end{equation}
Substituting this representation into matrix multiplication gives
\begin{equation}
y_{m,j}
=
\sum_{k=0}^{K_b-1}2^k
\left(\sum_i s_{m,i}^{(k)} b_{i,j}\right).
\label{eq:bit_plane_matmul}
\end{equation}
Thus, bit-plane encoding is mathematically exact, but each bit-plane
contribution must be weighted by its power-of-two factor.

\begin{figure*}[t]
    \centering
\includegraphics[width=\textwidth]{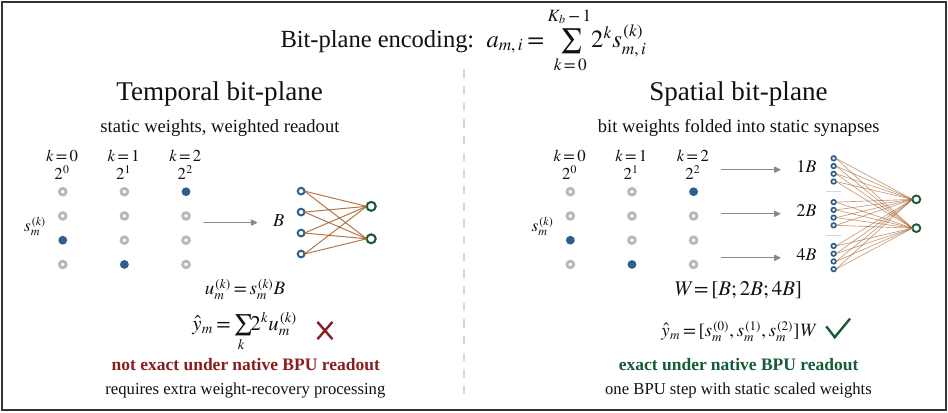}
\caption{Performance of bit-plane encoding in exact matrix multiplication equivalence}
    \label{fig:bit_plane_encoding_equivalence}
\end{figure*}

This weighting can be realized in two ways on BPU hardware, as illustrated in
Fig.~\ref{fig:bit_plane_encoding_equivalence}. In temporal bit-plane
execution, bit-planes are injected sequentially into the same BPU layer with
the unchanged synaptic matrix $\mathbf{B}$. The BPU computes unweighted
partial outputs, and the final weighting by $2^k$ must be performed at the
readout side. Therefore, temporal bit-plane encoding is mathematically exact
but not native direct-readout exact under a static-weight BPU execution model.

In spatial bit-plane encoding, different bit-planes are mapped to parallel
input groups and their weights are folded into static synapses:
\begin{equation}
w_{(k,i),j}=2^k b_{i,j}.
\end{equation}
The BPU accumulation then becomes
\begin{equation}
v_{m,j}
=
\sum_{k=0}^{K_b-1}\sum_i s_{m,i}^{(k)}w_{(k,i),j}
=
y_{m,j}.
\label{eq:spatial_bitplane_equivalence}
\end{equation}
Therefore, spatial bit-plane encoding completes the weighted accumulation
inside the BPU and supports native direct readout, at the cost of additional
input neurons, synapses, and a larger static weight range.

\subsection{Summary of BPU-Exact Equivalence and Spatio-Temporal Parallelism}\label{EE}

The above analysis shows that mathematical exactness and BPU-exact execution
are not the same. A spike encoding is mathematically exact if each activation
can be represented as a weighted sum of binary spike events. However, direct
exact output on real BPU hardware further requires that all contribution
weights be realized inside the BPU execution path. Under binary spike inputs,
static synaptic weights, and direct final-state readout, this can be achieved
in two cases:

\begin{itemize}
    \item \textbf{Native spike-count accumulation:} unit-weight spikes make
    the final neuron state directly equal to the matrix multiplication result.

    \item \textbf{Spatially folded weighted accumulation:} non-unit weights are folded into spatially expanded static synapses before deployment.
\end{itemize}

The former corresponds to temporal, spatial, and spatio-temporal spike-count encodings, while the latter corresponds to weighted encodings such as spatial bit-plane encoding. Temporal bit-plane encoding is mathematically exact, but its bit weights are not embedded in the static synaptic structure; therefore, it still requires readout-side weighted reconstruction and is not native direct-readout exact on the BPU.

This condition also exposes the spatio-temporal design space of exact
encodings. Temporal expansion saves hardware resources but increases latency,
whereas spatial expansion reduces time steps at the cost of more neurons,
synapses, and weight range. Spatio-temporal encoding balances these two
directions. Therefore, exact BPU execution depends on the joint design of
encoding, spatial mapping, synaptic weight scaling, and state readout,
motivating the BPU-SpikeMM mapping and execution framework in the next section.

\section{BPU-SpikeMM: Spatio-Temporal Mapping and Execution Framework}

\begin{figure*}[t]
    \centering
\includegraphics[width=\textwidth]{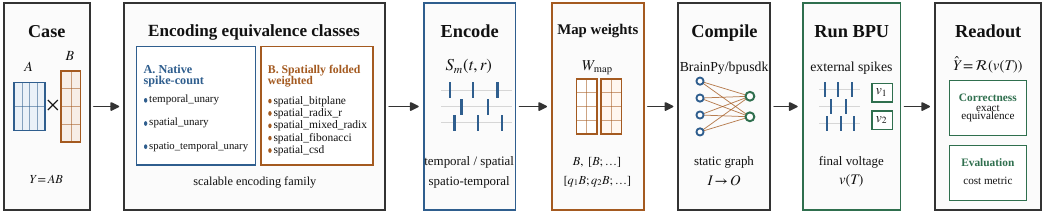}
\caption{Overview of the BPU-SpikeMM spatio-temporal mapping and execution framework}
    \label{fig:bpuspikemm_framework}
\end{figure*}

Based on the hardware-aware equivalence conditions derived above,
BPU-SpikeMM turns matrix multiplication into an executable BPU workload. The
framework takes a conventional multiplication case
$\mathbf{Y}=\mathbf{A}\mathbf{B}$ as input and produces two hardware-side
objects: a binary external spike workload derived from $\mathbf{A}$ and a
static synaptic weight block derived from $\mathbf{B}$. After compilation and
execution on real BPU hardware, the final membrane states of the output
neurons are read out for correctness checking and hardware-oriented metric
collection. The overall workflow is shown in
Fig.~\ref{fig:bpuspikemm_framework}.

The framework begins with row-wise problem decomposition. For a matrix
multiplication case, the activation matrix
$\mathbf{A}\in\mathbb{Z}_{\ge 0}^{M\times K}$ is processed row by row, while
the weight matrix $\mathbf{B}\in\mathbb{R}^{K\times N}$ is treated as the base
synaptic matrix. Each input row corresponds to one external spike workload,
and the $N$ output dimensions correspond to $N$ output neurons. The CPU result
$\mathbf{Y}=\mathbf{A}\mathbf{B}$ is kept as the reference for correctness
verification.

Next, BPU-SpikeMM selects an exact equivalence class. As summarized in the
left-middle block of Fig.~\ref{fig:bpuspikemm_framework}, supported
direct-readout mappings fall into two classes. The first is
\emph{native spike-count equivalence}, where values are represented by spike
counts and all spikes share the same unit contribution. The corresponding
synaptic block is therefore either $\mathbf{B}$ itself or repeated copies of
$\mathbf{B}$. The second is \emph{spatially folded weighted equivalence},
where non-unit contribution factors are folded into spatially expanded static
synaptic blocks before deployment. In both cases, the BPU receives only binary
spikes during execution; the difference lies in whether numerical scaling is
represented by spike count or by spatially scaled synapses.

Given the selected encoding, the encode stage converts each input row
$\mathbf{a}_m$ into a spatio-temporal spike representation $s_m(t,r)$, where
$t$ denotes the execution time step and $r$ denotes the spatial expansion
group. Temporal unary keeps the original input dimension and expands values
over time. Spatial unary moves this expansion into parallel input groups and
can finish in one time step. Spatio-temporal unary distributes the expansion
across both dimensions. Weighted encodings use one-step spatial expansion, in
which each spatial group corresponds to one encoded basis term.

In parallel, the map-weight stage constructs the static synaptic block
$\mathbf{W}_{\mathrm{map}}$. For native spike-count mappings,
$\mathbf{W}_{\mathrm{map}}$ is either $\mathbf{B}$ or a vertical repetition of
$\mathbf{B}$. For spatially folded weighted mappings, it is formed by stacking
scaled copies of the base matrix,
$\mathbf{W}_{\mathrm{map}}=[q_1\mathbf{B};q_2\mathbf{B};\ldots]$, where
$q_r$ is the contribution factor of the $r$-th spatial group. This step is the
key mechanism that turns mathematical weighted equivalence into
BPU-compatible direct-readout equivalence under static synaptic weights.

The compile stage builds a two-population BPU network with input and output
neurons. The mapped matrix $\mathbf{W}_{\mathrm{map}}$ is deployed as a dense
event-driven projection from the input population to the output population.
Output neurons are configured in a near-linear accumulation regime so that
their membrane states record accumulated synaptic contributions. The compiled
network and generated external spikes are then submitted to the BPU runtime.

During execution, the BPU receives the external binary spike stream and
performs event-driven synaptic accumulation. Since all contribution weights
required by direct-readout exact mappings have already been represented by
spike count or static spatial weight folding, no dynamic synaptic update is
needed. For the $m$-th input row, the final output state can be written as
\begin{equation}
    \mathbf{v}_m(T_m)=\sum_{t=1}^{T_m}\mathbf{s}_m(t)
    \mathbf{W}_{\mathrm{map}},
\end{equation}
which corresponds to the desired row output when the selected encoding
satisfies the equivalence condition.

Finally, the readout stage extracts the final hardware membrane voltage and
converts it to the integer-domain matrix multiplication result using the BPU
runtime scaling factor. The framework reports correctness against the CPU
reference, maximum absolute error, and hardware-oriented metrics, including temporal resource cost, spatial resource cost, and event-driven activity metrics. Therefore, BPU-SpikeMM serves not only
as an execution pipeline for exact matrix multiplication on real BPU hardware,
but also as a unified evaluation framework for spatio-temporal encoding and
weight-mapping strategies.

\section{Experimental Evaluation on Real BPU Hardware}
This section evaluates BPU-SpikeMM on physical BPU hardware with two
objectives. First, we verify BPU-native direct-readout exactness across
eight encoding schemes. Second, we characterize their trade-offs in
temporal cost, spatial resource consumption, event activity, and BPU
execution time. All runtime results are interpreted as intra-platform
comparisons among BPU mapping strategies under the same hardware and
software environment. Cross-architecture performance comparison with
conventional hardware is beyond the scope of this study.

\subsection{Experimental Environment}

The experiments were conducted on a BPU-based neuromorphic computing server. As shown in Table~\ref{tab:experimental_environment}, the software environment was built on Ubuntu 22.04.5 LTS with Python 3.10.12, NumPy 1.24.3, BrainPy 2.6.0.post20240803, and the BPU-SDK runtime. The hardware platform was equipped with 16 LYRA LRBM T276 BPU devices. Each experiment was executed on a single BPU; therefore, the reported results do not involve inter-BPU communication. Each BPU contains 108 tiles arranged in a $6 \times 18$ topology, and each tile integrates 4 NPUs. Each NPU provides 4096 neuron slots with a population group granularity of 1024 slots. Therefore, each BPU provides 1,769,472 theoretical neuron slots, and the full server provides 28,311,552 theoretical neuron slots across all 16 BPUs.

\begin{table}[t]
\centering
\caption{Experimental Environment}
\label{tab:experimental_environment}
\renewcommand{\arraystretch}{1.02}
\footnotesize
\begin{tabularx}{\columnwidth}{@{}p{0.45\columnwidth}>{\raggedright\arraybackslash}X@{}}
\toprule
\textbf{Category} & \textbf{Configuration} \\
\midrule
\multicolumn{2}{@{}l}{\textbf{Software Environment}} \\
Operating system & Ubuntu 22.04.5 LTS \\
Kernel version & Linux 5.15.0-163-generic \\
Programming language & Python 3.10.12 \\
Numerical computing library & NumPy 1.24.3 \\
SNN modeling framework & BrainPy 2.6.0.post20240803 \\
BPU SDK & BPU-SDK with BrainPyLib interfaces \\
BPU runtime & bpurun\_file\_multi\_stream \\
\midrule
\multicolumn{2}{@{}l}{\textbf{Hardware Environment}} \\
Number of BPUs & 16 (1 used per experiment) \\
BPU model & LYRA LRBM T276 \\
Tiles per BPU & 108 ($6 \times 18$) \\
NPUs per tile & 4 \\
Neuron slots per NPU & 4096 \\
Population group size & 1024 slots \\
Theoretical slots per BPU & 1,769,472 \\
Theoretical slots per server & 28,311,552 \\
\bottomrule
\end{tabularx}
\end{table}

\subsection{Correctness Verification}

Correctness in this work is defined as direct-readout exact equivalence on
real BPU hardware. Specifically, the matrix multiplication result is required
to be read directly from the final output neuron states after BPU execution,
without relying on weight recovery, bit-weight reconstruction, or host-side
weighted accumulation. In BPU-SpikeMM, direct-readout exactness is achieved by
either native spike counting or folding encoded spike contributions into
static synaptic weights before deployment. This property is important for
inference workloads, where deployed weights are reused across many input
activations. Based on this criterion, we verified the proposed spatio-temporal mappings for
temporal unary (T-Unary), spatial unary (S-Unary), spatio-temporal unary
(ST-Unary), spatial bit-plane (BitPlane), spatial radix-r (Radix-r), spatial mixed-radix (MixRadix), spatial Fibonacci (Fibo), and spatial CSD (CSD) encodings. Across all evaluated matrix sizes and encoding schemes, the BPU outputs exactly matched the CPU reference results, yielding a maximum absolute error of zero. These abbreviations are used consistently in the following
experimental figures.

\subsection{Experimental Analysis}

\begin{figure*}[t]
    \centering
\includegraphics[width=\textwidth]{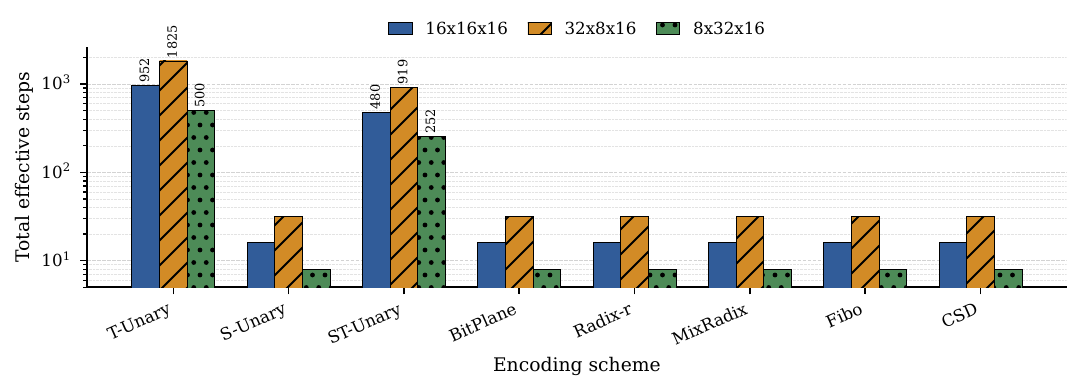}
\caption{Comparison of total effective time steps under different exact
spatio-temporal encoding schemes}
    \label{fig:total_effective_steps}
\end{figure*}

\begin{figure*}[t]
    \centering
\includegraphics[width=\textwidth]{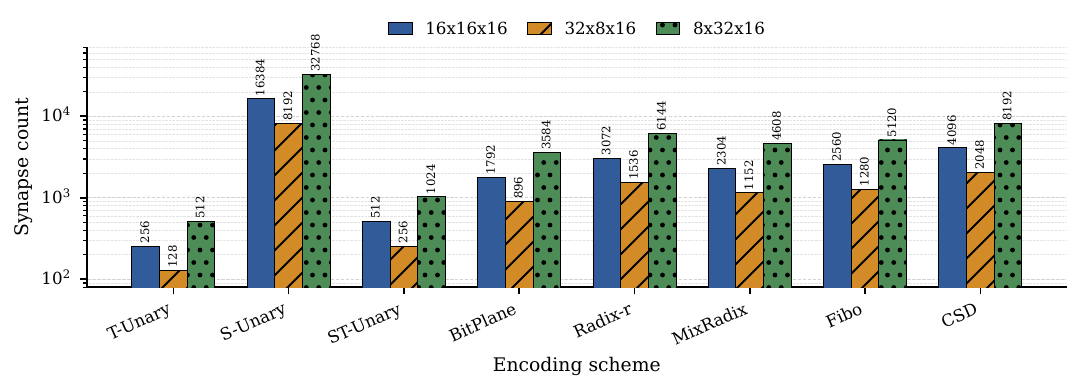}
\caption{Synapse count comparison under different exact encoding schemes}
    \label{fig:synapse_count}
\end{figure*}

\begin{figure*}[t]
    \centering
\includegraphics[width=\textwidth]{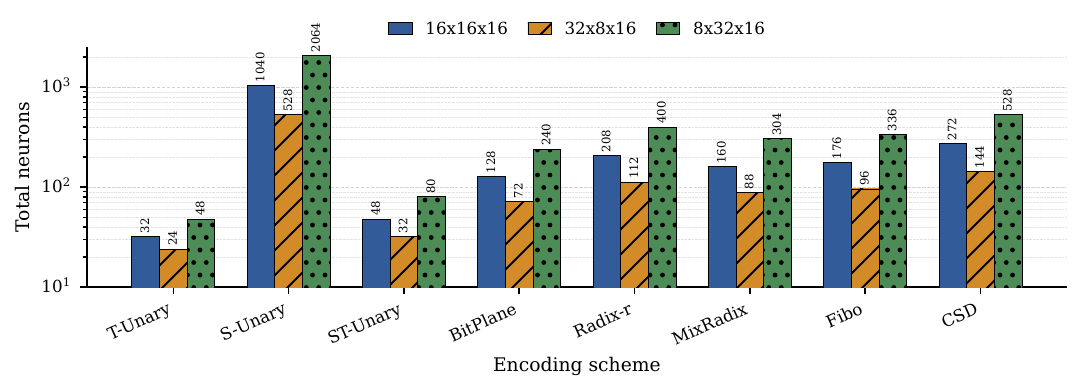}
\caption{Neuron count comparison under different exact encoding schemes}
    \label{fig:total_neurons}
\end{figure*}

Matrix multiplication can usually be decomposed into multiple smaller matrix
multiplication kernels through tiling. Therefore, this paper does not perform
an exhaustive evaluation over a large number of matrix sizes, but instead
selects representative basic computation scales for evaluation. Following the
tile-level execution style commonly used in matrix compute units such as
Tensor Cores, we consider three typical matrix multiplication sizes:
$16\times16\times16$, $32\times8\times16$, and $8\times32\times16$.
Here, each triplet denotes the $M$, $K$, and $N$ dimensions in
$\mathbf{A}\in\mathbb{R}^{M\times K}$ and
$\mathbf{B}\in\mathbb{R}^{K\times N}$. For each size, the input matrices are
randomly generated with integer values in the range $[0,64]$. Based on these
three representative scales, we evaluate eight exact spatio-temporal encoding
schemes under a unified setting, and analyze their differences in time-step
overhead, neuron count, synapse count, and related resource metrics.

As shown in Fig.~\ref{fig:total_effective_steps}, different encoding schemes
exhibit clear differences in temporal overhead. Temporal unary encoding
requires the largest number of effective time steps because each activation
value is represented by repeated spikes over time. Spatio-temporal unary
encoding reduces this overhead by moving part of the spike-count expansion to
the spatial dimension. In contrast, the fully spatial encodings complete each
input row in one BPU execution step, leading to much lower effective time-step
counts across all three tile sizes.

Fig.~\ref{fig:synapse_count} and Fig.~\ref{fig:total_neurons} report the
corresponding spatial resource costs. The results show that reducing temporal
overhead usually requires additional spatial resources. Spatial unary encoding
has the highest synapse and neuron counts because it expands spike counts
directly into replicated input lanes. Bit-plane, radix-based, Fibonacci, and
CSD encodings significantly reduce this spatial overhead by representing large
activation values with fewer active terms, while still folding their
contribution weights into static synaptic structures for direct BPU readout.

\begin{figure*}[t]
    \centering
\includegraphics[width=\textwidth]{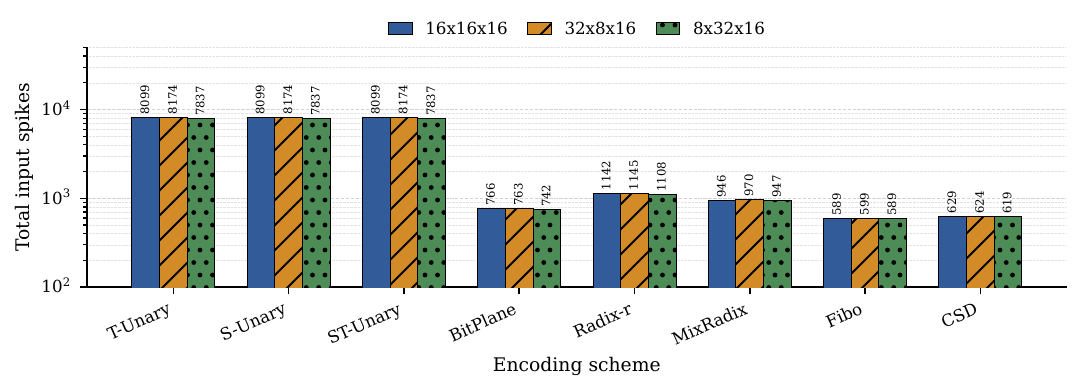}
\caption{Total input spike count under different exact encoding schemes}
    \label{fig:total_input_spikes}
\end{figure*}

\begin{figure*}[t]
    \centering
\includegraphics[width=\textwidth]{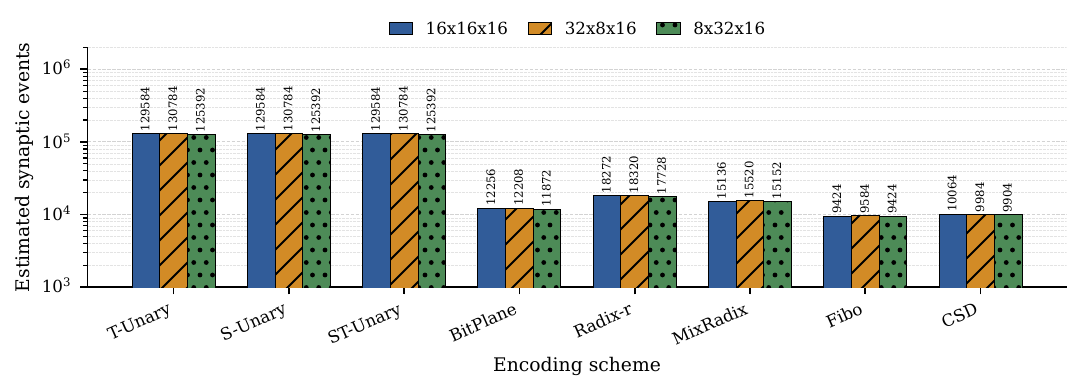}
\caption{Estimated synaptic events under different exact encoding schemes}
    \label{fig:synaptic_events}
\end{figure*}

The event-level behavior is further shown in Fig.~\ref{fig:total_input_spikes}
and Fig.~\ref{fig:synaptic_events}. Unary-based encodings produce more input
spikes because the activation magnitude is represented by spike count. In
comparison, weighted encodings reduce the number of active spike terms and
therefore lead to fewer estimated synaptic events. Among them, Fibonacci
encoding and CSD encoding achieve the lowest event counts in most cases,
showing their potential for reducing spike activity through more compact
number representations. This result suggests that exact encoding design is not
only a correctness problem, but also an important opportunity for reducing
event-level workload on real BPU hardware.

Overall, the results show clear trade-offs in three dimensions. In time,
fully spatial encodings are the fastest because each row is completed in one
BPU step, while temporal unary has the highest time-step overhead. In spatial
resources, temporal unary is the most compact because it reuses the original
weight matrix, whereas spatial unary consumes the most neurons and synapses.
In event activity, compact weighted encodings, especially Fibonacci and CSD,
usually generate fewer input spikes and synaptic events.

\begin{figure*}[t]
    \centering
\includegraphics[width=\textwidth]{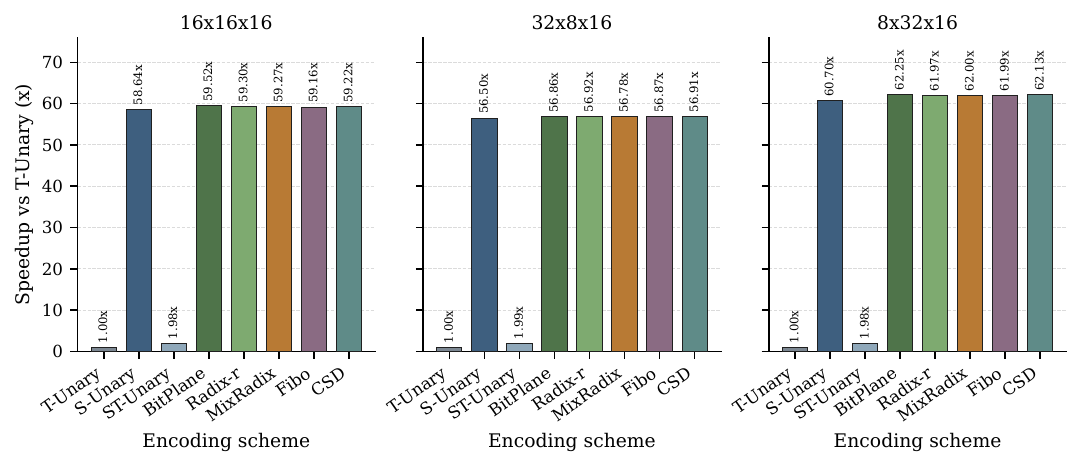}
\caption{Measured BPU execution speedup of different exact encoding schemes relative to temporal unary encoding}
    \label{fig:speedup_vs_temporal_unary}
\end{figure*}

To further examine the impact of different encoding schemes on measured BPU execution time, Fig.~\ref{fig:speedup_vs_temporal_unary} reports the
measured BPU runtime speedup of each encoding scheme, using temporal unary
encoding as the baseline for each tile size. The results are
consistent with the time-step analysis in Fig.~\ref{fig:total_effective_steps}.
Spatio-temporal unary encoding achieves about $2\times$ speedup by partially
moving the unary expansion from time to space. In contrast, fully spatial
encodings achieve much larger acceleration, typically around $56\times$ to
$62\times$ across the three tile shapes, because each row can be completed with
only one BPU execution step. Among the weighted spatial encodings, bit-plane,
radix-based, mixed-radix, Fibonacci, and CSD representations all maintain
similar high speedups, indicating that compact exact number representations can
substantially reduce resource and event overhead without sacrificing execution
latency. These measured results confirm that temporal expansion is the dominant
source of runtime overhead on BPU hardware, and that shifting computation into
the spatial dimension is an effective way to improve latency.

These results show that different exact encodings have different advantages,
depending on whether latency, hardware resources, or event activity is the
main concern. BPU-SpikeMM provides a unified framework to expose these
trade-offs on real BPU hardware. The code will be fully open-sourced to help
users evaluate suitable encoding strategies for their own scenarios.

\section{Conclusion}

This paper presented BPU-SpikeMM, an exact spatio-temporal spike encoding framework for matrix multiplication on real BPU neuromorphic hardware. By distinguishing mathematical equivalence from direct-readout exact equivalence, we established hardware-compatible mapping principles and built a unified pipeline from spike encoding and synaptic weight mapping to BPU execution and final-state readout. Experiments on real BPU hardware confirmed exact direct-output execution for eight spatio-temporal encoding schemes and revealed their trade-offs in temporal cost, spatial resources, and event activity. These results show that BPU-SpikeMM provides a practical basis for selecting suitable spike-domain matrix multiplication mappings and for supporting future hardware--software co-design. Future work will focus on migrating INT8-quantized large-model inference workloads to neuromorphic hardware.

\backmatter
%
% \bmhead{Acknowledgements}
% <confirmed acknowledgement text>
%
\section*{Declarations}

\textbf{Author contributions} Rongyuan Guo contributed to conceptualization,
methodology, software development, investigation, data analysis, visualization,
and writing the original draft. Yufeng Zhang contributed to conceptualization,
methodological guidance, project administration, and writing, reviewing, and
editing the manuscript. Haotian Wang contributed to methodology, experimental
validation, and data analysis. Canming Yao provided hardware platform support
and contributed to experimental deployment and validation. Yuxiang Huan
contributed to hardware environment configuration, experimental implementation,
and validation. Wangdong Yang contributed to theoretical analysis,
interpretation of the results, and manuscript review. Kenli Li contributed to
research supervision, project guidance, and manuscript review. All authors
reviewed and approved the final manuscript.

% References are generated in author--year style and alphabetized by
% the Springer Nature sn-basic bibliography style.
\bibliography{sn-bibliography}

\begin{thebibliography}{22}
\providecommand{\natexlab}[1]{#1}
\providecommand{\url}[1]{{#1}}
\providecommand{\urlprefix}{URL }
\providecommand{\doi}[1]{\url{https://doi.org/#1}}
\providecommand{\eprint}[2][]{\url{#2}}
 \bibcommenthead

\bibitem[{Bhattacharjee et~al.(2024)Bhattacharjee, Yin, Moitra, and Panda}]{bhattacharjee2024snns}
Bhattacharjee A, Yin R, Moitra A, et~al (2024) Are {SNNs} truly energy-efficient? a hardware perspective. In: Proceedings of the IEEE International Conference on Acoustics, Speech and Signal Processing (ICASSP), pp 13311--13315, \doi{10.1109/ICASSP48485.2024.10448269}

\bibitem[{Date et~al.(2023)Date, Kulkarni, Young, Schuman, Potok, and Vetter}]{date2023encoding}
Date P, Kulkarni S, Young A, et~al (2023) Encoding integers and rationals on neuromorphic computers using virtual neuron. Scientific Reports 13:10975. \doi{10.1038/s41598-023-35005-x}

\bibitem[{Demirkiran et~al.(2024)Demirkiran, Nair, Bunandar, and Joshi}]{demirkiran2024blueprint}
Demirkiran C, Nair L, Bunandar D, et~al (2024) A blueprint for precise and fault-tolerant analog neural networks. Nature Communications 15:5098. \doi{10.1038/s41467-024-49324-8}

\bibitem[{Jouppi et~al.(2023)Jouppi, Kurian, Li, Ma, Nagarajan, Nai, Patil, Subramanian, Swing, Towles, Young, Zhou, Zhou, and Patterson}]{jouppi2023tpuv4}
Jouppi NP, Kurian G, Li S, et~al (2023) {TPU} v4: An optically reconfigurable supercomputer for machine learning with hardware support for embeddings. In: Proceedings of the 50th Annual International Symposium on Computer Architecture (ISCA), pp 1--15, \doi{10.1145/3579371.3589350}

\bibitem[{Kim et~al.(2023)Kim, Kim, Um, Kim, Lee, and Yoo}]{kim2023snpu}
Kim S, Kim S, Um S, et~al (2023) {SNPU}: An energy-efficient spike domain deep-neural-network processor with two-step spike encoding and shift-and-accumulation unit. IEEE Journal of Solid-State Circuits 58(10):2812--2825. \doi{10.1109/JSSC.2023.3270442}

\bibitem[{Kudithipudi et~al.(2025)Kudithipudi, Schuman, Vineyard et~al.}]{kudithipudi2025scale}
Kudithipudi D, Schuman C, Vineyard CM, et~al (2025) Neuromorphic computing at scale. Nature 637(8047):801--812. \doi{10.1038/s41586-024-08253-8}

\bibitem[{Le~Gallo et~al.(2024)Le~Gallo, Hrynkevych, Kersting, Karunaratne, Vasilopoulos, Khaddam-Aljameh, Sarwat, and Sebastian}]{legallo2024aimc}
Le~Gallo M, Hrynkevych O, Kersting B, et~al (2024) Demonstration of 4-quadrant analog in-memory matrix multiplication in a single modulation. npj Unconventional Computing 1:11. \doi{10.1038/s44335-024-00010-4}

\bibitem[{Lei et~al.(2023)Lei, Flich, and Quintana-Ort{\'i}}]{lei2023toward}
Lei J, Flich J, Quintana-Ort{\'i} ES (2023) Toward matrix multiplication for deep learning inference on the {Xilinx Versal}. In: Proceedings of the 31st Euromicro International Conference on Parallel, Distributed and Network-Based Processing (PDP), pp 227--234, \doi{10.1109/PDP59025.2023.00043}

\bibitem[{Lu et~al.(2024)Lu, Long, Chen, Han, and Sun}]{lu2024aces}
Lu X, Long B, Chen X, et~al (2024) {ACES}: Accelerating sparse matrix multiplication with adaptive execution flow and concurrency-aware cache optimizations. In: Proceedings of the 29th ACM International Conference on Architectural Support for Programming Languages and Operating Systems (ASPLOS), pp 1--16, \doi{10.1145/3620666.3651381}

\bibitem[{Ma et~al.(2020)Ma, Sun, Qian, Meng, Yang, and Jiang}]{ma2020gounary}
Ma C, Sun X, Qian W, et~al (2020) Go unary: A novel synapse coding and mapping scheme for reliable {ReRAM}-based neuromorphic computing. In: Proceedings of the Design, Automation and Test in Europe Conference and Exhibition (DATE), pp 1--6, \doi{10.23919/DATE48585.2020.9116555}

\bibitem[{Ottati et~al.(2023)Ottati, Gao, Chen, Brignone, Casu, Eshraghian, and Lavagno}]{ottati2023spike}
Ottati F, Gao C, Chen Q, et~al (2023) To spike or not to spike: A digital hardware perspective on deep learning acceleration. IEEE Journal on Emerging and Selected Topics in Circuits and Systems 13(4):1015--1025. \doi{10.1109/JETCAS.2023.3330432}

\bibitem[{Parekh et~al.(2018)Parekh, Phillips, James, and Aimone}]{parekh2018threshold}
Parekh OD, Phillips CA, James CD, et~al (2018) Constant-depth and subcubic-size threshold circuits for matrix multiplication. In: Proceedings of the 30th ACM Symposium on Parallelism in Algorithms and Architectures, pp 67--76, \doi{10.1145/3210377.3210410}

\bibitem[{Pedersen et~al.(2024)Pedersen, Abreu, Jobst et~al.}]{pedersen2024nir}
Pedersen JE, Abreu S, Jobst M, et~al (2024) Neuromorphic intermediate representation: A unified instruction set for interoperable brain-inspired computing. Nature Communications 15:8122. \doi{10.1038/s41467-024-52259-9}

\bibitem[{Qin et~al.(2020)Qin, Samajdar, Kwon, Nadella, Srinivasan, Das, Kaul, and Krishna}]{qin2020sigma}
Qin E, Samajdar A, Kwon H, et~al (2020) {SIGMA}: A sparse and irregular {GEMM} accelerator with flexible interconnects for {DNN} training. In: Proceedings of the IEEE International Symposium on High Performance Computer Architecture (HPCA), pp 58--70, \doi{10.1109/HPCA47549.2020.00015}

\bibitem[{Schuman et~al.(2021)Schuman, Kay, Date, Kannan, Sao, and Potok}]{schuman2021sparse}
Schuman CD, Kay B, Date P, et~al (2021) Sparse binary matrix--vector multiplication on neuromorphic computers. In: 2021 IEEE International Parallel and Distributed Processing Symposium Workshops (IPDPSW), pp 308--311, \doi{10.1109/IPDPSW52791.2021.00054}

\bibitem[{Shrestha et~al.(2024)Shrestha, Timcheck, Frady, Campos-Macias, and Davies}]{shrestha2024efficient}
Shrestha SB, Timcheck J, Frady P, et~al (2024) Efficient video and audio processing with {Loihi 2}. In: Proceedings of the IEEE International Conference on Acoustics, Speech and Signal Processing (ICASSP), pp 13481--13485, \doi{10.1109/ICASSP48485.2024.10448003}

\bibitem[{Taka et~al.(2024)Taka, Gourounas, Gerstlauer, Marculescu, and Arora}]{taka2024gemm}
Taka E, Gourounas D, Gerstlauer A, et~al (2024) Efficient approaches for {GEMM} acceleration on leading {AI}-optimized {FPGA}s. In: Proceedings of the 32nd IEEE Annual International Symposium on Field-Programmable Custom Computing Machines (FCCM), pp 54--65, \doi{10.1109/FCCM60383.2024.00015}

\bibitem[{Theilman and Aimone(2025)}]{theilman2025fem}
Theilman BH, Aimone JB (2025) Solving sparse finite element problems on neuromorphic hardware. Nature Machine Intelligence 7:1845--1857. \doi{10.1038/s42256-025-01143-2}

\bibitem[{Vellaisamy et~al.(2023)Vellaisamy, Nair, Finn, Trivedi, Chen, Li, Lin, Wang, Blanton, and Shen}]{vellaisamy2023tubgemm}
Vellaisamy P, Nair H, Finn J, et~al (2023) {tubGEMM}: Energy-efficient and sparsity-effective temporal-unary-binary based matrix multiply unit. In: Proceedings of the IEEE Computer Society Annual Symposium on VLSI (ISVLSI), pp 1--6, \doi{10.1109/ISVLSI59464.2023.10238524}

\bibitem[{Wu et~al.(2020)Wu, Li, Yin, Hsiao, Kim, and San~Miguel}]{wu2020ugemm}
Wu D, Li J, Yin R, et~al (2020) {uGEMM}: Unary computing architecture for {GEMM} applications. In: Proceedings of the 47th Annual International Symposium on Computer Architecture (ISCA), pp 377--390, \doi{10.1109/ISCA45697.2020.00040}

\bibitem[{Yang et~al.(2024)Yang, Emer, and Sanchez}]{yang2024trapezoid}
Yang Y, Emer JS, Sanchez D (2024) Trapezoid: A versatile accelerator for dense and sparse matrix multiplications. In: Proceedings of the 51st Annual International Symposium on Computer Architecture (ISCA), pp 1--15, \doi{10.1109/ISCA59077.2024.00072}

\bibitem[{Yik et~al.(2025)Yik, Van~den Berghe, den Blanken et~al.}]{yik2025neurobench}
Yik J, Van~den Berghe K, den Blanken D, et~al (2025) The {NeuroBench} framework for benchmarking neuromorphic computing algorithms and systems. Nature Communications 16:1545. \doi{10.1038/s41467-025-56739-4}

\end{thebibliography}

\end{document}